\documentstyle[12pt]{article}

\def \sect #1 {\setcounter{equation} 0\section{#1}}
\def \be  {\begin{equation}}
\def \ee  {\end{equation}}
\def \ba  {\begin{eqnarray}}
\def \ea  {\end{eqnarray}}
\def \baa {\begin{eqnarray*}}
\def \eaa {\end{eqnarray*}}
\def \bb  {\begin{thebibliography}}
\def \eb  {\end{thebibliography}}
\def \cusp {{\rm cusp}}
\newcommand \ci [1] {\cite{#1}}
\newcommand \bi [1] {\bibitem{#1}}
\def \lab #1 {\label{#1}}
\newcommand\re[1]{(\ref{#1})}

\def \qqqquad {\qquad\qquad}
\newcommand\lr[1]{{\left({#1}\right)}}

\newcommand \ket [1] {|{#1}\rangle}
\newcommand \bra [1] {\langle {#1}|}
\def \e {\mbox{e}}
\def \CO {{\cal O}}

\def \CF {{\cal F}}
\def \W {\Sigma}
\def \PT {{\rm PT}}

\def \fracs #1#2 {\mbox{\small $\frac{#1}{#2}$}}

\def \bin #1#2 {{\left({#1}\atop{#2}\right)}}
\def \as {\relax\ifmmode\alpha_s\else{$\alpha_s${ }}\fi}
\def \alpi {\frac \as \pi}
\def \al #1 {\frac {\as({#1})}{\pi} }
\def \ds #1 {\ooalign{$\hfil/\hfil$\crcr$#1$}}
\def \MS {\overline{\rm MS}}

\def \tomega {W}
\def \d {{\rm d}}
\def \bi {\bibitem}
\def \CO {{\cal O}}

\begin{document}

\thispagestyle{empty}
\hfill\parbox{35mm}{{\sc ITP--SB--94--50}\par
                         hep-ph/9411211  \par
                         November, 1994}

\vspace*{30mm}
\begin{center}
{\LARGE Nonperturbative Corrections
 \\[3mm] in Resummed Cross Sections}
\par\vspace*{20mm}\par
{\large Gregory~P.~Korchemsky} and {\large George Sterman}

\par\bigskip\par\medskip

{\em Institute for Theoretical Physics,

State University of New York at Stony Brook,

Stony Brook, New York 11794 -- 3840, U.S.A.}
\end{center}
\vspace*{15mm}

\begin{abstract}
We show that the resummation of large perturbative corrections
in QCD leads to ambiguities in high energy cross sections
that are suppressed by powers of large
momentum scales. These ambiguities are caused by infrared
renormalons, which are a general feature of resummed hard-scattering
functions in perturbative QCD, even though these functions are infrared
safe order-by-order in perturbation theory. As in the case of the
operator product expansion, the contributions of infrared
renormalons to coefficient functions may be absorbed into the
definition of higher-dimensional operators, which induce nonperturbative
corrections that are power-suppressed at high energies. The strength of the
suppression is determined by the location of the dominant
infrared renormalon, which may be identified explicitly in the
resummed series. In contrast to the operator product expansion,
however, the relevant operators in factorized hadron-hadron
scattering and jet cross sections are generally nonlocal in QCD,
although they may be expressed as local operators in an effective theory
for eikonalized quarks. In this context, we verify and interpret the
presence of $1/Q$ corrections to the inclusive Drell-Yan cross
section with $Q$ the pair mass.
In a similar manner, we find $\exp(-b^2\ln Q)$ corrections
in the impact parameter space of the transverse momentum distributions of
 the Drell-Yan process and $\e^+\e^-$ annihilation.
We also show that the dominant nonperturbative corrections to cone-based
jet cross sections behave as $1/(Q\delta)$, with $\delta$ the opening
angle of the jet and $Q$ the center of mass energy.
\end{abstract}
\newpage

\section{Introduction}

It is almost axiomatic that
perturbation theory alone cannot give a full description of QCD.
Luckily, however, perturbative calculations may be supplemented
by nonperturbative functions and parameters to derive
physical predictions. While other sources of
nonperturbative behavior are possible, it is
attractive to identify those required to make perturbation theory
well defined. In this manner, we may use perturbation theory
as a diagonstic tool to identify a minimal set of nonperturbative
parameters.
This is the viewpoint that we shall advocate
and exploit below.
It is of particular interest for infrared safe quantities, which are
finite order-by-order in perturbation theory.  For such quantities,
we expect perturbation theory to predict leading behavior as
an asymptotic expansion in the coupling,
and for the failure of perturbation theory
at high orders to be particularly enlightening.

To be specific, we shall work in the class of quantities
for which
leading logarithmic behavior in energy
is described by Sudakov
resummation \cite{muc,cs,sen,korrad,kor}.
This resummation may be formulated in terms
of path-ordered exponentials, or Wilson lines.
Path-ordered exponentials have been extensively studied in
perturbation theory \cite{Polwil,korrad}
 and are also of special interest
for a nonperturbative description of the theory.  We will see
below that they naturally relate so-called infrared
renormalons \cite{irn1,mu1,irn2} to nonlocal operator expectation
values \cite{radm} for
cross sections of lepton pair production \cite{css,dy}
and energy-energy correlations \cite{cs} in ${\rm e}^+{\rm e}^-$
annihilation.  Generalizations
to other
quantities, including jet cross sections, and to
 nonleading logarithms are also discussed.

Already in \cite{ConS} it was shown that resummation
of large perturbative
corrections to the inclusive lepton pair cross section
leads to an ambiguity in perturbation theory
at the level of $\Lambda/Q$, with $Q$ the pair mass and $\Lambda$ the QCD
scale,
compared to leading power.
This paper is in large part an attempt to further understand
such power-suppressed effects\footnote{For recent
related studies see Ref.~\cite{manwise}.}.
We shall identify the corrections implied by these ambiguities
with specific nonlocal operators below.  In a similar manner,
we shall find
$\exp(-b^2\ln Q)$ corrections
(previously-discussed in \cite{cs})
in the impact parameter space
for the transverse momentum distribution
in the Drell-Yan cross section and in $\e^+\e^-$ annihilation.
In addition, we shall see that nonperturbative
corrections to cone-based jet cross sections
may be expected to begin at $1/(Q \delta)$,
with $\delta$ the opening angle of the jet and $Q$ the center of mass
energy. An application to inclusive B meson decay has been given
in \ci{us}.

Unaided, perturbation theory
fails in a number of ways.
At low orders,
infrared divergences and long-distance dependence
at fixed orders in perturbation theory
can be factorized
into nonperturbative functions like parton distributions, which are
 determined by experiment \cite{cssrv}.
In addition, QCD perturbation
theory is (probably) at best an asymptotic expansion
for Green functions, even off the mass shell.
Relevant for us are
 individual diagrams that behave at
$n$ loops as $\alpha_s^n n!$.  A series that behaves
in this fashion is not Borel summable,
let alone convergent, because
such terms produce
singularities in the Borel
transform that prevent it from being
inverted in a unique fashion.  Singularities that
are associated with individual diagrams are often termed
infrared or ultraviolet renormalons.
In QCD, it is the infrared (IR) renormalons that directly interfere with
the inversion of the Borel transform \cite{irn1,mu1}.

A lack of Borel summability does not
imply that a series is meaningless, however.  Rather,
it may represent
an asymptotic expansion for
a set of functions
that differ only in ways that do not affect this
expansion.  For instance, they may differ in powers
of $\exp[-1/\alpha]$, with $\alpha$ the expansion
parameter.  Different members of this
class of functions correspond to various definitions of the
inverse Borel transform.
It is this sort of ambiguity with which we shall
deal below, showing how its resolution requires the
introduction of new, nonperturbative, parameters associated
with vacuum expectations for
nonlocal operators\footnote{An application of similar ideas to QED is given in
\cite{DiCP}.}.

The paper is organized as follows. In Sect.~2 we  review how
infrared renormalons arise in resummed cross sections. Analyzing their
contributions to a Wilson line expectation value, we then find a general
form for power suppressed corrections
associated with infrared renormalons in the transverse
momentum distribution of lepton pairs. In Sect.~3
we relate the contribution of the leading infrared renormalon
to the matrix element of a nonlocal gluon field operator,
which can also be represented as a local operator in an effective field
theory for eikonal quarks.
In Sects.~4 and 5 we consider power corrections to jet cross sections and
inclusive lepton pair production, and identify
in each case a leading infrared renormalon
that results in $\Lambda/Q$ corrections. Sect.~6 contains concluding remarks.

\section{Infrared renormalons in a resummed cross section}

In many applications of perturbative QCD to high energy scattering, it
is desirable to sum finite corrections that remain after the
cancellation of infrared divergences. Examples are found in
energy-energy correlations in ${\rm e}^+{\rm e}^-$
annihilation \cite{cs,bbelbe}, the transverse momentum distribution of
lepton pairs \cite{earlydy,css}, Sudakov effects in elastic scattering
\cite{sudel}, the total cross section for lepton pair production
\cite{St87,CC,ConS1,kormar,ConS} and for heavy quark production \cite{LSvN},
and also in heavy quark effective theory \cite{hqet,us}.
As we shall see, such
resummations of perturbative corrections generally imply the presence of
nonperturbative corrections suppressed by powers of kinematic invariants.

Let us first consider the transverse momentum distribution of
lepton pairs in the Drell-Yan process, $h+ h'\to \ell^+\ell^-+X$.
The relevant formalism is virtually identical for energy-energy correlations
in ${\rm e}^+{\rm e}^-$ annihilation,
closely related to jet structure \cite{cs}.
For large transverse momentum of the lepton pair, $Q_t\sim Q$,
the differential cross-section for the process
may be written to leading power in the lepton invariant mass $Q$ as
\begin{equation}
{\d \sigma_{hh'}\over \d Q^2 \d Q_t^2} = \sigma_0 \W(Q^2,{ Q}_t^2)\,,
\lab{QTsimQ1}
\end{equation}
where $\sigma_0={4\pi^2\alpha^2}/{9Q^2s}$
is the inclusive Born cross
section for a quark of unit charge, where
$s$ is the squared invariant mass of the incoming hadrons,
and where the function $\W$ depends on parton distributions
$f_{a/h}(x,\mu^2)$ through the factorization formula \cite{cssrv},
\begin{equation}
\W(Q^2,{ Q}_t^2)
= \int_0^1 {\d x_a \over x_a} {\d x_b \over x_b} f_{a/h}(x_a,Q^2)
f_{b/h'}(x_b,Q^2)
H_{ab}\biggl ( {Q^2\over x_a x_b s},{Q_t^2\over Q^2} \biggr )\, .
\lab{QTsimQ2}
\end{equation}
The bulk of the lepton-pair cross section, however, is in the region
$Q_t^2\ll Q^2$. In this region the function
$\W$ contains ``double-logarithmic" corrections like
 $1/Q_t^2 \times \alpha_s^n \ln^{2n-1}(Q_t^2/Q^2)$.
The resummation of these large corrections is
most easily expressed
in the Fourier ${\bf b}$-space conjugate to ${\bf Q}_t$,
\be
\W(Q^2,{ Q}_t^2)=\int {\d^2b \over (2\pi)^2}\
\e^{i{\bf Q}_t\cdot{\bf b}}\ {\tilde\W(Q^2,b^2)}\, ,
\label{sigfortrans}
\ee
where $b^2\equiv {\bf b}^2$.
Then, the double-logarithmic corrections to ${\tilde\W(Q^2,b^2)}$
are of the form $\alpha_s^n \ln^{2n}  Q^2b^2$ and exponentiate.
At leading power in $Q^2$, the expression for ${\tilde\W(Q^2,b^2)}$
is given by \cite{cs,css}
\be
{\tilde\W(Q^2,b^2)}=
\int_0^1 {\d x_a \over x_a} {\d x_b \over x_b} f_{a/h}(x_a,C/b^2)
f_{b/h'}(x_b,C/b^2)\delta(1-Q^2/x_ax_bs)\ \e^{-S(Q^2,b^2)}\, ,
\lab{wq2}
\ee
where $\e^{-S(Q^2,b^2)}$ is a resummation factor that includes the large
logarithms in $Q^2b^2$,
\begin{equation}
S(Q^2,b^2)
=\int_{C/b^2}^{Q^2} {\d k_t^2 \over k_t^2 }
\left[ \ln{Q^2\over k_t^2}
\Gamma_{\cusp}(\alpha_s(k_t^2)) + \Gamma(\alpha_s(k_t^2))\;
\right]\,,\quad\quad
C=4\e^{-2\gamma}\,,
\lab{Wresum}
\end{equation}
with $\gamma$ the Euler constant.
Here,
$\Gamma_{\cusp}(\alpha_s)$ is the universal ``cusp" anomalous dimension
\ci{korrad,kor}, while
$\Gamma(\as)$, which contributes to nonleading behavior in
$\ln Q$, depends on
the specific process and choice of the constant $C$.
With the choice above,
both anomalous dimensions begin at order $\alpha_s$,
\begin{equation}
\Gamma_{\cusp}(\as)=\alpi C_F+O(\alpha_s^2)\, ,
\quad\quad
\Gamma(\as) = -{3\over 2} \alpi C_F+O(\alpha_s^2)\, ,
\lab{gamc}
\end{equation}
with $C_F=4/3$.
A short calculation, using the one-loop form of the running coupling,
\begin{equation}
\alpha_s(k_t^2)={1\over \beta_1 \ln(k_t^2/\Lambda^2)}\,,\qqqquad
\beta_1=(11-2n_f/3)/(4\pi)\, ,
\lab{alphas}
\end{equation}
shows that the leading (double)
logarithms of $Q^2b^2$
found order-by-order are replaced in the resummed form
by
\be
S(Q^2,b^2)
=\frac{C_F}{\pi\beta_1}
\left[
\ln(Q^2/\Lambda^2)
\ln\frac{\ln(Q^2/\Lambda^2)}{\ln(C/b^2\Lambda^2)}
-\ln(Q^2b^2/C)
\right]
\,,
\lab{dl}
\ee
where we have used the one-loop expression \re{gamc} for $\Gamma_{\cusp}$ and
have neglected nonleading logarithmic terms due to $\Gamma$ in \re{Wresum}.
At large $b$, the Fourier transform
${\tilde\W(Q^2,b^2)}$ is strongly suppressed by the resummed
exponent, which makes the structure function $\W(Q^2,{ Q}_t^2)$
amenable to perturbative treatment.
We also note, however, that the exponent diverges at $b^2=C/\Lambda^2$, because
of the singularity in the running coupling (\ref{alphas})
at $k_t^2=\Lambda^2$. It is the
consequences of this singularity for power corrections in $b^2$ that we
shall discuss below. First of all, we notice that approaching the ``critical''
value of the impact parameter $b\sim 1/\Lambda$, one has to take into account
power corrections $\CO(\Lambda^2 b^2)$, which are negligible with respect to
logarithmic corrections $\ln(\Lambda^2 b^2)$ for small $b^2$, but which
become
important for large $b^2$. We stress that the factorization in \re{wq2} was
found in the leading $1/Q^2$ limit and therefore in principle includes
power corrections of $\CO(\Lambda^2 b^2)$ provided that $b^2 Q^2\gg 1$, or
equivalently $Q_t^2\ll Q^2$.
At the same time, the expression \re{Wresum} was obtained
by summing all logarithmic corrections and neglecting power corrections.
Thus, to study the $b^2$ power corrections we have to find an approximation
for the exponent
$S(Q^2,b^2)$ that includes both logarithmic and power corrections in $b^2$.
To this end, we recall the origin of
$\e^{-S(Q^2,b^2)}$ in the factorized expression
\re{wq2}.

In the Drell-Yan process, the resummation factor, $\e^{-S(Q^2,b^2)}$,
appears as the
contribution of real and virtual soft gluons interacting with the quark and
antiquark that annihilate to produce a virtual photon. The
interaction of soft gluons with quarks can be treated using the eikonal
approximation and can be summarized in terms of path-ordered exponentials
(Wilson lines) \cite{kormar} as follows. Let us introduce a notation
for ``straight" path-ordered exponentials in the direction of momentum $p$,
\be
\Phi_p[Sp+x,S_0p+x]={\rm P} \exp \bigg ( -ig\int_{S_0}^S \d s\, p^\mu
A_\mu(ps+x) \bigg )\, ,
\label{phioexp}
\ee
where P denotes ordering in group indices of the gluon field $A_\mu(x)$.
Similarly,
for products of two such ordered exponentials, defined by the light-like
momenta $p_1$ and $p_2$ of the incoming quark and antiquark,
 respectively, we define
\be
U_{\rm DY}(x) = {\rm T} \bigg [\,
\Phi_{p_2}^\dagger(x,-\infty)\;
\Phi_{p_1}(x,-\infty)
\bigg ]= {\rm T} \bigg [\,
\Phi_{-p_2}(-\infty,x)\;
\Phi_{p_1}(x,-\infty)
\bigg ]
\, , \qquad p_1^2=p_2^2=0\,,
\lab{phidef}
\ee
where T is time-ordering. Finally, in terms of such path-ordered
exponentials,
we define the eikonalised Drell-Yan cross-section as \cite{kormar}
\begin{equation}
\tomega_{\rm DY}(b,\mu) =
\langle 0|\;U^\dagger_{\rm DY}(0)\; U_{\rm DY}(b)\;|0\rangle\, ,
\qquad
b^\nu=(0,0,{\bf b})
\lab{tomegadef}
\end{equation}
with $b_\nu b^\nu\equiv -b^2$.
If we insert a complete set of states, $1=\sum_X\ket{X}\bra{X}$
between the operators $U$ and $U^\dagger$,
\be
\tomega_{\rm DY}(b,\mu)=\sum_X\left|\bra{0} \; U_{\rm DY}(0)\;\ket{X}\right|^2
\exp(-i{\bf k}_t\cdot{\bf b})\,,
\ee
the Wilson line expectation value $\tomega_{\rm DY}$ describes
the cross section for the emission of gluons of any energy by
two eikonalized quarks, weighted
by a factor $\exp(-i{\bf k}_t\cdot{\bf b})$, with ${\bf k}_t$ the
total transverse momentum of gluons in the final state $\ket{X}$. Such a cross
section is ultraviolet divergent, and requires renormalization, which makes
$\tomega_{\rm DY}$ a function of renormalization scale
$\mu$. 
As discussed in \ci{St87}, for instance,
$\tomega_{\rm DY}$  takes into account the contribution
of soft gluons to the Drell-Yan cross section.
The value of $\mu$ is arbitrary but the most convenient
choice is $\mu\sim Q$.
Let us show that for $\mu=Q$ the expression \re{tomegadef} indeed
coincides with the resummation factor \re{dl}
up to power corrections in $b^2$.

Expanding
\re{tomegadef} in powers of the gauge field,
we get the following unrenormalized expression for $\tomega_{\rm DY}$
to the lowest order of perturbation theory:
\be
W^{(1)}_{\rm DY}=1+g^2\mu^{4-D}C_F\int\frac{\d^Dk}
{(2\pi)^D}2\pi\delta_+(k^2)
\lr{\frac{p_1}{p_1\cdot k}-\frac{p_2}{p_2\cdot k}}^2
(1-\e^{-i{\bf k}_t\cdot{\bf b}})\,,
\label{lowestWDY}
\ee
where $\mu$ is the parameter of dimensional regularization,
with $D=4-2\varepsilon$,
$\d^Dk\equiv \d k^+\d k^-\d^{D-2}k_t$, and ${\bf k}_t$ is the
$(D-2)-$dimensional
transverse momentum of the gluon. For massless quarks, the $k^\pm$ integral
contains a collinear divergence.
This divergence is eliminated, however, when
one evaluates the logarithmic derivative ${\d\ln W_{\rm DY}}/{\d\ln Q^2}$.
Using the one-loop expression $W_{\rm DY}^{(1)}$ and the
exponentiation of infrared divergences in
hard-scattering processes \ci{korrad,kor,gsgath} we get
\be
\frac{\d\ln W_{\rm DY}}{\d\ln Q^2}
=4C_F\mu^{4-D}\int\frac{\d^{D-2}k_t}{(2\pi)^{D-2}}
\frac{\as}{ k_t^2}\lr{\e^{-i{\bf k}_t\cdot{\bf b}}-1} \,.
\lab{diff}
\ee
To take into account higher order corrections to \re{diff}
we choose the argument of the coupling constant to be
the transverse momentum, $\as=\as(k_t^2)$, and identify
$C_F\alpha_s/\pi$ as the lowest term in the expansion of the cusp anomalous
dimension, $\Gamma_{\cusp}(\as(k_t^2))$ \cite{korrad,kor},
\be
\frac{\d\ln W_{\rm DY}}{\d\ln Q^2}
=
4\pi \mu^{2\varepsilon}\int \frac{\d^{2-2\varepsilon}k_t}
{(2\pi)^{2-2\varepsilon}}
\frac{\Gamma_{\cusp}(\alpha_s(k_t^2))}{k_t^2}
\bigg ( \e^{-i{\bf k}_t\cdot{\bf b}}-1 \bigg )\, .
\lab{DYevol}
\ee
Notice that after integration over large $k_t^2$ this expression contains
ultraviolet poles in $\varepsilon$ which need to be subtracted.
The resulting renormalized expression contains both logarithmic
and power corrections in $b^2$ to the eikonalized cross section, due to
soft gluon emissions from the incoming quarks. Neglecting power corrections to
\re{DYevol}, we find by comparison with eq.~\re{Wresum}, that
the behavior of $\ln\tomega_{\rm DY}$ in $\ln Q$ in \re{DYevol} is the same
as the behavior of $S(Q,b)$ in $\ln Q$ up to nonleading
logarithmic terms due to the ``collinear'' anomalous dimension $\Gamma$
in \re{Wresum}.
Identification of the resummation factor $\exp(-S(Q^2,b^2))$ as a
Wilson line expectation value \re{DYevol} allows us now to analyze the
perturbative contribution of soft gluons with small transverse momentum
$k_t^2$ to the Drell-Yan cross section.

In the evolution equation \re{DYevol} and its solution $W_{\rm DY}$
we find divergences at $k_t^2=\Lambda^2$. Notice that \re{DYevol} was
found after resummation of soft gluons to all
orders of perturbation theory,
and that this singularity can easily be translated into the
high-order behavior of the original perturbative series before resummation.
We can reconstruct the expansion in $\alpha_s(Q^2)$ by using the one
(or higher) loop running coupling,
$\alpha_s(k_t^2)=\alpha_s(Q^2)/[1-\beta_1\alpha_s(Q^2)\ln Q^2/k_t^2]$.
Such an expansion generates a sum of integrals of the form
$\int_0^1 \d x \ln^n(1/x)=\Gamma(n+1)$, and we find that the singularity
manifests itself in the perturbative expansion in $\alpha_s(Q^2)$ as an
infrared renormalon in the exponent \cite{ConS}, behaving at $n$th loop
order as $\alpha_s^n \beta_1^n n!$.

Let us now study in more detail
how IR renormalons appear in the evolution
equation \re{diff}, including renormalization.
After substitution of the relation
$\alpha_s(k_t^2)
=\int_0^\infty d\sigma \lr{{k_t^2}/{\Lambda^2}}^{-\sigma\beta_1}$
into \re{diff}, and after integration over transverse momenta, we get
\begin{equation}
\frac{\d\ln W_{\rm DY}}{\d\ln Q^2}=-\frac{C_F}{\pi}(\pi\mu^2 b^2)^\varepsilon
\int_0^\infty
\frac{\d\sigma}{\sigma\beta_1+\varepsilon}
\frac{\Gamma(1-\sigma\beta_1-\varepsilon)}{\Gamma(1+\sigma\beta_1)}
\bigg ({\frac{\Lambda^2{ b}^2}{4}}\biggr )^{\sigma\beta_1}\, .
\lab{15}
\end{equation}
We notice
that $(\Lambda^2 b^2/4)^{\sigma\beta_1}
={\rm e}^{-\sigma/\alpha_s}$ with $\alpha_s=\alpha_s(4/ b^2)$
and identify the right-hand side of the last relation
as the Borel representation,
\be
\pi(\as)=\int_0^\infty \d\sigma \tilde\pi(\sigma)\ {\rm
e}^{-\sigma/\alpha_s}\,,
\ee
of
$\pi(\alpha_s)\equiv{\d\ln W_{\rm DY}}/{\d\ln Q^2}$, with
\begin{equation}
\tilde\pi(\sigma)=-\frac{C_F}{\pi}(\pi\mu^2{ b}^2)^\varepsilon
\frac1{\sigma\beta_1+\varepsilon}
\frac{\Gamma(1-\sigma\beta_1-\varepsilon)}{\Gamma(1+\sigma\beta_1)}\, .
\lab{16}
\end{equation}
Let us analyze the singularities of the Borel transform $\tilde\pi(\sigma)$
for real positive $\sigma$.
In the limit $\varepsilon\to 0$ the function
$\tilde\pi(\sigma)$ has a pole for $\sigma=0$. Since small values of the Borel
parameter correspond to small coupling constants $\alpha_s(k_t^2)$, or
equivalently to large transverse momenta $k_t^2$, the singularity
at $\sigma=0$ is
ultraviolet.
Away from $\sigma=0$ we
put $\varepsilon=0$ and find that the function $\tilde\pi(\sigma)$ has
singularities generated by $\Gamma-$functions at
$
\sigma^*=1/\beta_1,\ 2/\beta_1,\ 3/\beta_1,\ \ldots \, ,
$
the infrared renormalons. To define the integral over large $\sigma$
we have to fix the prescription for integration of the IR renormalons in
\re{15}.
Different prescriptions lead to results that differ from each other in
powers of $b^2\Lambda^2$, beginning at
$\CO(b^2\Lambda^2)$, the contribution
of the ``leading'' IR renormalon, at $\sigma^*=1/\beta_1$.

To get a ``perturbative''
approximation to $W_{\rm DY}$,
we treat $\sigma$ as a small parameter
and then integrate \re{15} over
$0<\sigma<1/2\beta_1$.
For small $\sigma$ we replace the ratio of $\Gamma-$functions in \re{15} by
$\e^{2\gamma\sigma\beta_1+\gamma\varepsilon}$,
use the expansion $1/(\varepsilon+\sigma\beta_1)=
1/\varepsilon-\sigma\beta_1/\varepsilon^2+\ldots$
and change variables to $x=\sigma\beta_1$ to get
\be
\frac{\d\ln W_{\PT}}{\d\ln Q^2}=\frac{C_F}{\pi\beta_1}
\sum_{n=1}^\infty (-)^n
\frac{(2\pi\mu^2b^2/C)^\varepsilon}{\varepsilon^n}
\int_0^{1/2}\d x\;x^{n-1}(\Lambda^2{ b}^2/C)^x\,,
\ee
with $C$ given in eq.~(\ref{Wresum}).
Subtracting poles in the $\MS$ scheme, we find after summation
the following result, for $\mu^2=2 Q^2$ and $1/{ b}^2\gg\Lambda^2$,
\be
\frac{\d\ln W_{\PT}}{\d\ln Q^2}=
\frac{C_F}{\pi\beta_1}
\int_{1/\ln(Q^2/\Lambda^2)}^{1/\ln(C/\Lambda^2 b^2)}
\frac{\d x}{x}+{\cal O}(1)
=
\frac{C_F}{\pi\beta_1}
\ln\frac{\ln(C/\Lambda^2{ b}^2)}{\ln(Q^2/\Lambda^2)}
+{\cal O}(1)\, .
\lab{17}
\ee
One easily checks that the resummation factor \re{dl} satisfies this relation
up to nonleading logarithms in $Q$,
so that to this approximation, $W_{\PT}=\exp\left (-S(Q^2,b^2)\right )$.

There is no unique way of defining the $k_t$ integral  in
eq.~(\ref{DYevol}) or the $\sigma$ integral in eq.~\re{15}.
As discussed above, however, we may
consider the singularities of the running
coupling as inducing an {\it ambiguity\/} in the integral, which is
to be eliminated by adding new, nonperturbative parameters to the
theory \cite{mu1,mu2}. This process is not utterly arbitrary, because
the singularity appears only at small values of $k_t$, where, unless $b$ is
very large ({\it i.e.}, of order $1/\Lambda$), the integrand is suppressed.
This means that we may consistently define the full integral in the
right-hand side
of \re{DYevol} as a power series expansion in $b$, starting with a
``perturbative" contribution $S_\PT(b)=-\ln W_\PT(b)$, eq.~\re{17},
at order $(b^2)^0$:
\be
\e^{-S(Q^2,b^2)}\approx
W_{\rm DY}(b)
=\exp\left(-S_\PT(b)- b^2 S_2(Q) - b^4 S_4(Q) + \ldots\right )\,,
\lab{main}
\ee
where the equality is approximate,
 since we have neglected nonleading collinear
corrections to $S(Q^2,b^2)$. The fact that the Borel transform \re{16} has
IR renormalon singularities
at $\sigma^*=n/\beta_1$ for $n=1,2,...$ implies that
nonperturbative effects should contribute
$\CO(b^{2n}\Lambda^{2n})$ power corrections to the exponent in \re{main},
or equivalently to the functions $S_{2n}(Q)$. Moreover,
the explicit form of the evolution equation \re{DYevol} implies that in
the large $Q$ limit these functions behave as
%
\be
S_{2n}(Q) \sim A_{2n} \ln Q + B_{2n}\,.
\lab{ln}
\ee
This is because the IR renormalons arise from small values of $k_t$
in \re{DYevol}, where we may take
$\epsilon=0$. Then
the contribution of the IR renormalons in \re{DYevol} to
${\d W_{\rm DY}}/{\d Q^2}$
is independent of $Q$, which, along with \re{main}, implies the functional
dependence \re{ln}.

\section{Operator content of the leading IR renormalon}

We now identify the operator content of the leading infrared renormalon,
$\sigma^*=1/\beta_1$, which contributes to the $b^2\Lambda^2$ correction
to the
right-hand side
 of \re{DYevol}, by recognizing it as the first term in the
expansion of $\ln\tomega_{\rm DY}$ with respect to $b^2$. We denote this
contribution
as
\begin{equation}
S_2(Q)=- {\partial \over \partial b^2}
\ln \tomega_{\rm DY}(b)
\bigg |_{b=0} \,.
\lab{tildEop}
\end{equation}
Applying this derivative to the logarithm of
$\tomega_{\rm DY}$, eq.~(\ref{tomegadef}), evaluated at $b=0$, we find
\be
S_2(Q)=
{1\over 4}\bra{0}\; U_{DY}^\dagger(0)\;(i\vec\partial)^2\;
U_{DY}(0)\;\ket{0}\,,
\label{EopU}
\ee
where the derivative $\vec\partial=(\partial_1,\partial_2)$ acts in the
transverse
$b-$space. Finally, using the definition \re{phidef}
of $U_{\rm DY}$ we
get a form reminiscent of \re{lowestWDY},
\be
S_2(Q)= {1\over 4}\bra{0}
\Bigg|\Phi^\dagger_{p_2}(0,-\infty)
\lr{\vec\CF_{p_1}(0) - \vec\CF_{p_2}^\dagger(0)}
\Phi_{p_1}(0,-\infty)\Bigg|^2 \ket{0}\, ,
\lab{tildEop2}
\ee
where
${\cal F}_{p}^\alpha(x)\equiv
\lr{iD^\alpha\Phi_p(x,-\infty)} \Phi_p^\dagger(x,-\infty)$ is a
nonlocal ``eikonalized" field strength,
which may also be written in terms of
the local field strength $F^{\mu\nu}$ and
the ordered exponential $\Phi_p$, eq.~(\ref{phioexp}), as
\begin{equation}
{\cal F}_{p}^\alpha(x)
=-ig\int_{-\infty}^0 \d s\ \Phi_p(x,x+sp)\;p_\mu F^{\mu\alpha}(sp+x)\;
\Phi_{-p}(x+ps,x)\, .
\lab{Fdef}
\end{equation}
The perturbative contributions to $S_2$
are scaleless integrals and vanish in dimensional regularization.
This is similar to the perturbative renormalization of the
divergent matrix elements of a local operator such as
$\langle 0| F^2|0\rangle$.  Similarly, we may take
$\tomega_{\rm DY}(0)$ as unity.

{}From \re{tildEop2} and \re{Fdef}, we see that the leading
nonperturbative correction is described by
the matrix element of a nonlocal operator of the gauge field. This
is in contrast with the total $\e^+\e^-$ cross section, in which the leading
IR renormalon appears at $\sigma=2/\beta_1$ and gives rise to
power corrections described by the local operator $\langle 0| F^2|0\rangle$.
Nevertheless, it is possible to represent \re{tildEop2} as a
local operator in
an effective field theory.
In this model, the quark with light-cone
momentum  $p$ is described by the effective field, $q_p(x)$, and the
interaction with gauge fields, $A_\mu(x)$, is organized
to reproduce the quark-gluon interaction in the eikonal approximation.
The Lagrangian of the model is given by
\be
{\cal L}=\bar q_p(x) (p \cdot iD) q_p(x)\,,
\ee
where $\bar q_p=q_p^\dagger$, with $p$ the quark momentum, and where
$D_\mu=\partial_\mu - i gA_\mu(x)$ is the covariant derivative in the
quark representation.
 For a heavy quark with mass
$M$, velocity $v_\mu$ and momentum $p=Mv$, this is the Lagrangian
of heavy quark effective theory. The solution of the equation
of motion for
the $q-$field reproduces the eikonal phases of the quark fields
\be
q_p(x)=\Phi_{p}(x,-\infty)\; a_p\, ,
\ee
where $a_p=q_p(-\infty)$ is the creation operator of the eikonalized quark.
Using this nonabelian ``Bloch-Nordsieck" model, we may rewrite $U_{\rm DY}(x)$,
eq.\ (\ref{phidef}), as a composite local operator in the effective theory,
$
U_{\rm DY}(x)\sim \bar q_{p_2}(x) q_{p_1}(x),
$
where fields $q_{p_1}$ and $\bar q_{p_2}$ describe a quark and an antiquark
with momenta $p_1$ and $p_2$, respectively.
The fields $q$ carry color indices of the quarks, and the summation
over these indices is implied.
In these terms, the matrix element in \re{EopU} is given by
\be
E_2(Q) = {1\over 4}
\bra{0} (\bar q_{p_1} q_{p_2})
\;(i\vec\partial)^2\;
(\bar q_{p_2} q_{p_1})
\ket{0}\,,
\lab{E2}
\ee
provided that
$
\bra{0} a_{p_1}^\dagger a_{p_1}\ket{0}
= \bra{0} a_{p_2}^\dagger a_{p_2}\ket{0}= 1\,.
$
As was pointed out above, the value of this nonperturbative
matrix element will depend on how we define perturbation theory.
One definition is to apply a principal value
prescription for integration around the poles in the inverse
Borel transform, eq.~(\ref{15}) \cite{ConS}.

A different method
is implicit in the prescription given
 by Collins and Soper \cite{cs}, in which the impact parameter
$b$ is replaced by the function
\begin{equation}
b^*=\frac{b}{[1+b^2Q_0^2]^{1/2}}\,,\qquad Q_0\sim 2\ {\rm GeV}\, ,
\lab{bstar}
\end{equation}
in the expression for the exponent $S(Q^2,b^2)$
and for the quark distribution functions $f_{a/h}(x,C/b^2)$ in
eq.~\re{wq2},
while an additional
nonperturbative factor
\begin{equation}
\exp\bigg ( {-g_1(b^2,\tau)-g_2(b^2)\ln\frac{Q}{2Q_0}} \bigg )
\lab{3}
\end{equation}
is introduced into the exponent $S$ in eq.~\re{Wresum} to suppress the
large-$b$ region in the Fourier integral in eq.~(\ref{sigfortrans}).
In the large-$b$ limit, $b^*$ approaches a maximum value of $1/Q_0$ and
the perturbative
$k_t^2$ integral in \re{Wresum} becomes well-defined.
For moderate $1/Q\ll b \ll 1/\Lambda$, the following parameterization
was also proposed, again consistent with the result of
eq.~(\ref{main}) above,
\begin{equation}
g_1(b^2,\tau)=g_1 b^2 + {\cal O}(b^4),
\qquad
g_2(b^2)=g_2 b^2 + {\cal O}(b^4).
\lab{4}
\end{equation}
The values of $g_1$ and $g_2$ cannot be calculated in perturbative QCD,
but have been estimated \cite{dy} from experiment%
\footnote{Another parameterization was proposed in \ci{LY}.},
$$
g_1=0.15\ ({\rm GeV})^2, \quad \quad  g_2=0.40\ ({\rm GeV})^2\,.
$$
In summary, the results \re{main} and \re{ln} of our analysis are
in agreement with the Collins-Soper parameterization of nonperturbative
effects in \re{3} and \re{4}.
In addition, our considerations give a field-theoretic interpretation to
the coefficient $g_2$, as
\begin{equation}
g_2 =\frac {\d S_2(Q)}{\d \ln Q}=A_2
\lab{g2E}
\end{equation}
with $S_2$ and $A_2$ given in (\ref{tildEop2}) and \re{ln}.
Of course, the specific value of $g_2$ quoted above
depends on the Collins-Soper method, \re{bstar} and \re{3}, of defining
the perturbative content of resummation.  Other values will give
different $g_2$'s in general, an ambiguity that we expect from
general considerations.

It is now useful to discuss the nature of our approximations.
We have taken into account the contributions of soft gluon radiation
from active quarks, and have ignored the contributions of collinear gluons as
nonleading in $\ln Q^2$.
In perturbation theory, this property follows from the
expression \re{Wresum}
for the resummation factor, in which purely collinear interactions
contribute to the anomalous dimensions $\Gamma$, while soft emissions
contribute to both $\Gamma$ and $\Gamma_{\cusp}$.
Generalizing this observation to the power corrections in the
exponent $S$, eq.~\re{main},
we find that collinear interactions modify the
coefficients $B_{2n}$ in \re{ln},
but not the leading coefficients $A_{2n}$.
In addition, considering the expression \re{wq2} for $\tilde\W(Q^2,b^2)$, we
notice that for large $b^2$ the normalization scale in the definition
of the quark distribution functions becomes small,
which should produce
nonperturbative power corrections to $f_{a/h}(x_a,C/b^2)$.
Such corrections are naturally attributed to ``intrinsic"
motion of quarks inside hadrons. This motion affects the transverse momentum
of the lepton pair, and
may be parameterized phenomenologically by the
function $g_1(b^2,\tau)$ in \re{3}.
This implies that the coefficients $B_{2n}$ in \re{ln}
reflect the intrinsic
momenta of quarks in hadrons
\footnote{Similar phenomena occur in the behavior of the heavy quark
distribution function in the end-point region \cite{us}.}.
In realistic hadronic states, we expect quarks to have a ``residual''
virtuality, which should replace $1/b^2$ as a regulator of collinear
divergences
in $S(Q^2,b^2)$, \re{Wresum}, for large $b^2$. At the same time, the
coefficients $A_{2n}$ are free from collinear divergences, {\it i.e.},
are infrared safe, and are thus independent of
quark virtuality. Therefore, we expect the coefficients $A_{2n}$ to
be independent of the incoming hadrons.
This explains why only
the vacuum expectation values of operators appear
in relations \re{tomegadef} and \re{tildEop2}.

It is clear that this pattern will recur whenever a cross section
can be written in terms of a resummed expression like eq.~(\ref{Wresum}),
that involves the integral over the scale of the running coupling.
In the following two sections, we shall find two other applications,
jet cross sections and the inclusive dileption cross section, which
show a similar pattern, and which illustrate the
dependence of the implied nonperturative corrections on the process.

\section{Power corrections in jet cross sections}

As another application of our infrared renormalon analysis, we examine power
corrections to infrared-safe jet cross sections.
Let us consider
the cross section for ${\rm e}^+{\rm e}^-\to 2\ {\rm jets}$,
defined by calorimeters in the form of
back-to-back cones of half-angle $\delta$ in the overall center of
mass. Let
$\sqrt{s}=Q$ be the center of mass energy, and
let $xQ$ be the total energy
flowing into the two jets. For $x\to 1$, the cross-section
$\d\sigma_{2{\rm J}}/\d x$
receives large
perturbative corrections associated with
the emission of soft gluons outside the jet
cone, with total energy $(1-x)Q$.

As in the
Drell-Yan process,
the asymptotic behavior of the cross section
$\d\sigma_{2{\rm J}}/\d x$ for $x\to 1$
is best seen in terms of its Fourier transform
with respect to the total
energy $(1-x)Q$ carried by soft gluons,
\be
\frac{\d\sigma_{2{\rm J}}((1-x)Q,\delta)}{\d x}
=\int_{-\infty}^\infty \frac{\d y_0}{2\pi}\
\e^{-iy_0Q(1-x)}\; \tilde\sigma(y_0,\delta)\, .
\lab{tildesig}
\end{equation}
For large $y_0$ (conjugate to $x\to 1$), large perturbative corrections
to the Fourier transform $\tilde\sigma(y_0,\delta)$ exponentiate in
much the same way as for the transverse momentum distribution
in the Drell-Yan process
discussed above, due to
the factorization of soft gluons from the two energetic jets. As in the
previous case, we neglect the nonleading contributions of collinear gluons,
and concentrate on the leading behavior in $\delta$ only, given by
\be
{\tilde \sigma}(y_0,\delta)=\exp(-S_{\rm 2J}(y_0,\delta))\, .
\lab{s2J}
\ee
To lowest order the exponent
$S_{\rm 2J}$ is given by (compare \re{lowestWDY}),
\be
S_{\rm 2J}(y_0,\delta)=
g^2C_F\int\frac{\d^D k}{(2\pi)^D}\; \theta_{\rm jet}(k)\; 2\pi\delta_+(k^2)
\lr{\frac{p_1}{p_1\cdot k}-\frac{p_2}{p_2\cdot k}}^2
(1-\e^{ik_0 y_0})\, ,
\lab{sim}
\ee
with $p_1={Q\over 2}(1,1,{\bf 0})$ and $p_2={Q\over 2}(1,-1,{\bf 0})$
the light-like momenta of quarks in the
two energetic jets.
The function $\theta_{\rm jet}(k)$ is zero when $k$ is in a jet cone,
and unity otherwise.
Making explicit this kinematical restriction on the
momenta of soft
gluons, and including higher order corrections to
$S_{2{\rm J}}(y_0,\delta)$, we get,
to leading logarithmic accuracy,
\be
S_{\rm 2J}(y_0,\delta)
= -2\int_0^Q\frac{\d k_0}{k_0}\lr{1-\e^{ik_0y_0}}
  \int_{k_0^2\delta^2}^{k_0^2}\frac{\d k_t^2}{k_t^2}
  \Gamma_{\cusp}(\alpha_s(k_t^2))\,,
\lab{Edef}
\ee
where integration is performed over energy and over
transverse momentum with respect to the axes of the outgoing jets.
The soft gluon energy in \re{Edef} is restricted to be less than
the energy of the ``parent'' quark, while the lower limit,
$k_t >k_0\delta$,
implies that the soft gluon is emmitted outside the jet cone.
After
substitution of \re{Edef} and \re{s2J} into \re{tildesig} and integration
over $y_0$, eq.~\re{tildesig} reproduces the
leading perturbative series for
the
eikonalized 2-quark jet cross section with the total energy of gluons
outside the cones given by $(1-x)Q$.

To take into
account higher order corrections to the cross section, we have
chosen, as usual,
the argument
of the coupling constant to be $k_t^2$.
An evolution equation for the exponent, $S_{\rm 2J}(y_0,\delta)$, now
follows from \re{Edef}
(compare eq.~(\ref{DYevol})) in terms of the ``cone"
parameter $\delta$,
\be
\frac{{\rm d} S_{2{\rm J}}(y_0,\delta)}{{\rm d}\ln \delta}
=4\int_0^Q\frac{\d k_0}{k_0}\lr{1-\e^{ik_0y_0}}
\Gamma_{\cusp}(\alpha_s(k_0^2\delta^2))\, .
\lab{Eevol}
\ee
On the right-hand side of this equation,
an infrared renormalon arises
once again, from integrations over soft gluon energy
$k_0\sim\Lambda/\delta$, due to the singularity in
$\alpha_s(k_0^2\delta^2)$. Taking the one-loop expression \re{gamc} for
the cusp anomalous dimension and using 
the representation
$\alpha_s(k_0^2\delta^2)=\int_0^\infty d\sigma
(k_0^2\delta^2/\Lambda^2)^{-\sigma\beta_1}$, we may perform the
integration over $k_0$ in \re{Eevol} to get an expression similar to \re{15}.
However, an important difference from \re{15} is that the IR renormalons
now appear at $\sigma^*=1/2\beta_1, 1/\beta_1, 3/2\beta_1, ...$ . Thus, in
contrast with the transverse momentum distribution, the leading IR renormalon
appears in the $\e^+\e^-\to 2\ {\rm jets}$ cross section at
$\sigma^*=1/2\beta_1$, which gives rise to an $\CO(y_0\Lambda/\delta)$ power
correction to the exponent, $S_{2{\rm J}}(y_0,\delta)$ in \re{s2J}.
We conclude that in the jet cross section, $\tilde\sigma(y_0,
\delta)$, the leading nonperturbative power corrections occur at level
$y_0\Lambda/\delta$.

Now suppose we regulate the exponent $S_{\rm 2J}(y_0,\delta)$
by some procedure, for example, a $y_0$-dependent
cutoff, by analogy to the Collins-Soper procedure in \re{bstar} and \re{3},
or a
principal value prescription for integrating over IR renormalons.
We denote the result by $S_{{\rm 2J},\PT}(y_0,\delta)$. Taking into account the
contribution of the leading IR renormalon to $S_{\rm 2J}(y_0,\delta)$, we may
write the cross section as
\begin{equation}
{\tilde \sigma}(y_0,\delta)
=
\exp\lr{-S_{{\rm 2J},\PT}(y_0,\delta) - i\frac{y_0\Lambda}{\delta} A_{{\rm 2J}}
+\CO(y_0^2)}\;,
\lab{npsig2J}
\end{equation}
with $A_{{\rm 2J}}$ a new dimensionless nonpertubative parameter.
In place of eq.~(\ref{tildesig}), we now have
\begin{equation}
{\d \sigma_{2{\rm 2J}}(xQ,\delta) \over \d x}
=
\int_{-\infty}^\infty \frac{\d y_0}{2\pi}
\e^{-iy_0Q(1-x+A_{{\rm 2J}}
\Lambda/Q \delta)}\; \exp\lr{-S_{{\rm 2J},\PT}(y_0,\delta)}\, .
\lab{tildesigA}
\end{equation}
As usual, $A_{{\rm 2J}}$ is dependent on our definition of
$S_{{\rm 2J},\PT}$.
We thus see that the effect of the leading infrared renormalon is to shift
the scaling variable $x$ by an amount inversely proportional to the typical
transverse
momentum in the jets, $Q\delta$.

An operator intepretation of this result is not as simple as for
the examples discussed in the previous section, because eq.~(\ref{Edef})
requires restrictions on both angles and energies of the soft gluons
that give double logarithms.  Nevertheless, the same leading logarithmic
corrections can be generated by introducing
a product of outgoing massive
eikonal lines, defined as in eq.~(\ref{phidef}), but with
\begin{equation}
p_1^2=p_2^2=m^2\,,\qquad m^2=Q^2\delta^2 \, .
\lab{jetmass}
\end{equation}
The corresponding operator which takes into account soft gluon emissions into
the final state is now
\begin{equation}
U_{2{\rm J}}(x)={\rm T}
\bigg [ \Phi_{p_1}(\infty,x)\Phi_{p_2}^\dagger(\infty,x) \bigg ]\, ,
\lab{uepemdef}
\end{equation}
and a corresponding cross section for the emission of gluons
of total energy $(1-x)Q$ is given by \re{tildesig} with
\be
\tilde\sigma(y_0,\delta)
=
W_{\e^+\e^-}\equiv
\bra{0} U^\dagger_{2{\rm jet}}(0) U_{2{\rm jet}}(y)\ket{0}\, ,
\lab{tomegaepem}
\ee
where $y^\mu=(y_0,0,{\bf 0})$. Indeed, eq.\ \re{tomegaepem} has the same
perturbative expansion as \re{s2J} and \re{sim},
except that now the quarks have time-like momenta, $p_1=(E,p,{\bf 0})$ and
$p_2=(E,-p,{\bf 0})$ with $E^2=p^2+m^2$, and the gluon momentum integrals are
unrestricted. Then, integrating over $k_3$
 we get the following
leading logarithmic expression for the exponent $S_{\rm 2J}(y_0,\delta)$,
\be
S_{\rm 2J}(y_0,\delta)
= -2\int_0^\infty\frac{d k_0}{k_0}\lr{1-\e^{ik_0y_0}}
  \int_{k_0^2m^2/E^2}^{k_0^2}\frac{dk_t^2}{k_t^2}
  \Gamma_{\cusp}(\alpha_s(k_t^2))\, ,
\lab{mass}
\ee
where $E\approx Q/2$ is the quark energy. Due to the cusp singularities of the
Wilson lines, the $k_0-$integral in \re{mass} is ultraviolet divergent
and should be regularized, for example, dimensionally,
as in eq.\ \re{diff}. The net effect of the regularization is
to introduce a cut-off $\mu$ for large energies $k_0$. Taking $\mu=Q$ and
identifying the mass of the quark as in \re{jetmass}, we find that \re{mass}
coincides with expression \re{Edef} for $S_{\rm 2J}(y_0,\delta)$.
Clearly, to incorporate nonleading logarithms, which depend, in
particular, on the internal structure of the jet, we must go beyond this
rather crude approximation. Besides these nonleading terms, there are
additional corrections in this model associated with an ambiguity in the
definition of the mass of eikonalized quarks. As discussed in \ci{mass},
the mass itself suffers from an ultraviolet renormalon, which introduces an
ambiguity of order $\Lambda$ into its definition.
The corresponding ambiguity in the final-state energy contributes
a term of order $\Lambda y_0$ to the exponent
$\exp[-iy_0 Q(1-x+A_{2{\rm J}}\Lambda/Q\delta)]$
in eq.\re{tildesigA}, down by a factor of $\delta$ from the
contribution we have identified.

Since the first infrared renormalon singularity is
proportional to $y_0$, we may identify the parameter $A_{\rm 2J}$
with the first derivative of $\tomega_{{\rm e}^+{\rm e}^-}(y_0,\delta)$
with respect to $y_0$ at $y_0=0$, by analogy to eq.~(\ref{tildEop2})
above, and we find that
\begin{equation}
A_{\rm 2J}
= \langle 0|
{\rm \bar T}\bigg [
\Phi_{p_1}(0,\infty)\lr{
{\cal F}_{0p_1}(0)
       -
{\cal F}_{0p_2}^\dagger(0
       )}
\Phi^\dagger_{p_2}(0)
\bigg]^\dagger
{\rm T}\left[\Phi_{p_1}(0) \Phi^\dagger_{p_2}(0)\right]
|0\rangle\, .
\lab{tildEop3}
\end{equation}
 Note that this form involves only a single factor of the field
 strength, made gauge invariant by its combination with path-ordered
 exponentials.

One may be tempted to suggest that  such a matrix element, with
only a single field-strength, must vanish.  In fact, if
we were allowed to remove the time and anti-time ordering from
\re{tildEop3}, this would be the case. That is,
we could use the property $\Phi_{p}^\dagger \Phi_{p}=1$
of ordered exponentials to simplify \re{tildEop3}
to $\bra{0}\left ( {\cal F}_{0p_1}(0)-{\cal F}_{0p_2}^\dagger(0)
\right )\ket{0}$.
This matrix element, however,
 is zero, because $\langle 0| {\cal F}_{\mu,p}|0\rangle = c p_\mu$,
with $c$ some constant,
while $p^\mu\langle 0|{\cal F}_{\mu p}|0\rangle
=\langle 0|(p\cdot D)\Phi_p
\Phi_p^\dagger|0\rangle =0$ from the definition of the Wilson line,
which implies that $c=0$.
This argument fails, however, because the gauge fields
on the $p_1$ and $p_2$ eikonal lines do not commute in general.
Thus, the matrix element in \re{tildEop3} need not be zero.
As usual, its precise value will depend on how we
define the ambiguity in the resummed perturbation series.  The
important point here is that the matrix element depends
on a single, although nonlocal, vacuum matrix element.
A more detailed
 study of the large variety of jet cross sections will result
 in further information.

\section{Power corrections in inclusive lepton-pair production}

As final example, we consider the inclusive Drell-Yan cross section
$\d \sigma_{{\rm DY}}(\tau,Q^2)/\d Q^2$, normalized to a
structure function of deeply inelastic scattering,
$F_{{\rm DIS}}(x,Q^2)$
\cite{St87,CC,ConS1,kormar,ConS}.
To be specific,
the moments of the dilepton
hard-scattering function are related to those of the
DIS structure functions by
\begin{equation}
\int_0^1 \d\tau \tau^{n-1} {\d\sigma_{{\rm DY}}(\tau,Q^2)\over\d Q^2}
=\sigma_0\;
\omega(n,Q^2)\; F_{\rm DIS}^2(n,Q^2)\, ,
\end{equation}
where $\tau=Q^2/s$ and $\sigma_0$ is the inclusive
Born cross section, as in eq.~(\ref{QTsimQ1}), and where
$F_{\rm DIS}(n,Q^2)$ is the corresponding moment of
$F_{\rm DIS}(x,Q^2)$ with respect to $x$.
For large values of $n$ (conjugate to $\tau\to 1$
and $x\to 1$), $\omega(n,Q^2)$
contains large perturbative corrections
of the form $\alpha_s^k\ln^{2k}n$.

The logarithms of $n$, however, are readily resummed.  In fact,
up to corrections that are bounded in $n$, the moments of
the hard-scattering function exponentiate in the form
\begin{equation}
\omega(n,Q^2)
=
R(\alpha_s(Q^2))\,
\exp\left [E(n,Q^2)\right ]\, ,
\label{omegaE}
\end{equation}
where $R(\alpha_s)$ is a finite function of
$\alpha_s$, while
the function $E(n,Q^2)$ generates all logarithmic corrections in $n$
from the integrals\footnote{In \cite{CC} and \cite{ConS1}, $2\Gamma_{\cusp}$
and $2\Gamma$ are denoted $g_1$ and $g_2$, respectively.}
\begin{equation}
E(n,Q^2)
=
- 2 \int_0^1 \d z\; {z^{n-1}-1 \over 1-z}\; \left [
\int_{(1-z)^2Q^2}^{(1-z)Q^2} {\d k_t^2 \over k_t^2}\;
\Gamma_{\cusp}(\alpha_s(k_t^2))
+ \Gamma\left(\as((1-z)Q^2)\right)\, \right ]\, .
\label{Ekdef}
\end{equation}
The anomalous dimensions $\Gamma_{\cusp}$ and $\Gamma$
have been encountered above in eq.~(\ref{gamc}).
$E(n,Q^2)$ contains infrared renormalon singularities
from small values of $k_t$.
We may isolate the leading infrared renormalon by interchanging the
$k_t$ and $z$ integrals in the first term of (\ref{Ekdef}),
\begin{eqnarray}
E(n,Q^2)
&=&
- 2 \int_0^{Q^2} {\d k_t^2 \over k_t^2}\;
\Gamma_{\cusp}(\alpha_s(k_T^2))
\int_{1-k_t/Q}^{1-k_t^2/Q^2} \d z\;
 {\left (1-(1-z)\right )^{n-1} -1  \over 1-z} \nonumber \\
&=&
2(n-1)\int_0^{Q^2} {\d k_t^2 \over k_t^2}\;
\Gamma_{\cusp}(\alpha_s(k_t^2))
 \left [ {k_T\over Q} - {k_T^2\over Q^2}
 \left ({n-2\over 4}\right) \right ] + \cdots\, ,
\label{Eexpand}
\end{eqnarray}
where we have expanded $z^{n-1}$ in
powers of $1-z$, and have suppressed
$(1-z)^3$ and higher.  Such terms produce
higher powers of $nk_t/Q$.  For large values of $k_t$, all such
terms contribute to the perturbative expansion of $E(n,Q^2)$.  The
leading infrared renormalon, however, which may be identified in the same
manner as above in eq.~(\ref{15}), is present only in the
first term in the expansion of (\ref{Eexpand}), which we denote
\begin{equation}
{4n A_{\rm DY}(Q) \over Q} \equiv 4{n\over Q} \int_0^{Q^2} \d k_T\;
\Gamma_{\cusp}(\alpha_s(k_T^2))\, .
\end{equation}
Again, $A_{\rm DY}(Q)$ includes both perturbative and nonperturbative
contributions.  The latter is precisely the $1/Q$ contribution
identified in \cite{ConS}.
Following our procedure
above, we now find a nonlocal matrix element
which corresponds to $A_{\rm DY}(Q)$.

We begin by defining a product of Wilson lines that
reproduces the logarithmic behavior of the Drell-Yan cross
section near the boundary of phase space \ci{kormar},
\begin{equation}
W_{\rm DY}(k_0,\mu)= \int {\d y_0\over 2\pi}
\e^{-ik_0y_0}
\langle 0|U^\dagger_{\rm DY}(0)\; U_{\rm DY}(y)|0\rangle\; ,
\end{equation}
with $U_{\rm DY}$ the product of ordered exponentials in the directions
of the incoming particles, given above in eq.~(\ref{phidef}) for
$x=y\equiv(y_0,{\bf 0})$.
As indicated, $W_{\rm DY}$ requires renormalization at the scale $\mu$.
{}From the procedure introduced
in \cite{St87},
$W_{\rm DY}$ satisfies an evolution equation whose solution
is the convolution form\footnote{The reasoning is essentially identical to
that given in Section 5 of \cite{St87}.},
\begin{eqnarray}
W_{\rm DY}(k_0,\mu)
&=&
{\tilde R}(\alpha_s,\varepsilon)
\sum_{n=0}^\infty {2^n\over n!}\; \prod_{i=1}^n\;
\int_0^1 \d z_i
\Bigg [ {1\over 1-z_i}\int_0^{(1-z_i)\mu} {\d k_{t,i}^2 \over
k_{t,i}^{2+2\varepsilon}}\;
\Gamma_{\cusp}(\alpha_s(k_{t,i}^2))
\nonumber \\
&\ & \hbox{\hskip 1.25 true in}
+{\tilde \Gamma}\left(\as((1-z)\mu^2)\right)\; \Bigg ]_+
 \delta\left (k_0 -\sum_1^n(1-z_i)\mu\;
\right )\, .
\label{WDYconv}
\end{eqnarray}
Here, $\tilde R$ is an (infrared sensitive) function of $\alpha_s$
and $\varepsilon$, and $\tilde \Gamma$ organizes nonleading logarithms,
which may differ from those in $E(n,Q^2)$ above.
We may now evaluate the Fourier transform of $W_{\rm DY}$ explicitly,
\begin{equation}
\tilde{W}_{\rm DY}(y_0,\mu)
=
\int {\d k_0 \over 2\pi} e^{ik_0y_0}
W_{\rm DY}(k_0,\mu)\, .
\end{equation}
We then recognize that its derivative with respect to $y_0$,
evaluated at $y_0=0$, generates
the product of ${\tilde W}_{\rm DY}(0,\mu)$, which can be normalized
to unity, and a single integral over
$k_t$, in which appears in the scale of the running
coupling,
\begin{equation}
-i {\partial \ln \tilde{W}_{\rm DY}(y_0,\mu) \over \partial y_0}|_{y_0=0}
=
2\int_0^{\mu^2} {\d k_t^2 \over k_t^{2+2\varepsilon} }
\Gamma_c(\alpha_s(k_t^2))\;
(\mu-k_t)\, .
\label{firstderiv}
\end{equation}
While both of the terms on the right-hand side
contain an infrared renormalon, only the second corresponds
to the function $A_{\rm DY}(\mu)$ identified from the resummed cross section
above.
 To isolate
$A_{\rm DY}(\mu)$ from the ultraviolet
contribution in eq.~(\ref{firstderiv}),
we simply form the combination,
\begin{equation}
A_{\rm DY}(\mu)
=
-{i\over 2}\;
{\partial \ln \tilde{W}_{\rm DY}(y_0,\mu) \over \partial y_0}|_{y_0=0}
-
{1\over \mu}\;
{\partial^2 \ln \tilde{W}_{\rm DY}(y_0,\mu) \over \partial y_0^2}|_{y_0=0}\, .
\label{kfactop}
\end{equation}
Here, the second term, which is evaluated just as in (\ref{firstderiv}),
serves only to cancel the large, ultraviolet contribution.  Its
nonperturbative contribution, however, begins only
at order $\mu^{-2}\sim Q^{-2}$,
 which we neglect in our approximation.  Note that
the second term is not unique, and higher derivatives (times higher
powers of $1/\mu$) would equally well cancel the ultraviolet
contribution, while contributing infrared renormalons that
are yet further suppressed in $\mu\sim Q$.

\section{Interpretation}

We have studied several
cross sections, whose leading logarithms are
generated by products of ordered exponentials.  In each case, we found
that nonleading power corrections are required by the
presence of infrared renormalons in the corresponding
resummation formulas.  These corrections, in turn, may
be represented as vacuum matrix elements of
field strengths, integrated over the paths of the original
ordered exponentials.
It is important to stress that our results do not depend on
an extrapolation of the $k_t^2$ integral in,
for example, eq.\
(\ref{Wresum}), to soft momenta, $k_t^2\sim \Lambda^2$.  Rather,
eq.~(\ref{Wresum}) organizes the large-order behavior of the cross
section that follows from the evolution equation (\ref{DYevol}).
In this sense, the
presence of the divergence in the running coupling follows from
the factorial behavior of perturbation theory, not the other way
around.

In the specific cross sections we have studied, we have found that
the first nonleading-$Q$ behavior  is associated with
nonperturbative matrix elements, such as (\ref{tildEop2}),
(\ref{tildEop3}) and (\ref{kfactop}).
{}From our critereon of consistency for the
full theory, including both perturbative and nonperturbative
contributions, we conclude that these power corrections exponentiate, to
give, for example, a nonperturbative Gaussian distribution in $b$.
The Fourier transform of such a
function gives a Gaussian behavior in momentum space, which
must be convoluted with the perturbative $Q_T$ distribution.
On the other hand, behavior linear in
the transform variable, such as $n/Q$ in the Drell-Yan
normalization and $y_0/Q$ in the
jet cross section, is associated with a shift in the
conjugate kinematic variable.

The reasoning that leads to this picture
is similar to that described in \cite{mu1},
which connected infrared renormalons in the total ${\rm e}^+{\rm e}^-$
annihilation cross section with the local condensate
$\langle 0|F^2(0)|0\rangle$.  Here, we must go beyond the
class of local operators to ``nonlocal condensates", but with
the benefit of greater flexibility, and perhaps, generality
of application.  Of course,
this procedure lacks the guiding principles of
the operator product expansion, and we have chosen to let
perturbation theory suggest for us the form of
nonperturbative structures that we may expect.  While other
contributions, not related to perturbation theory, are
probably present,
the self-consistency of the theory demands the presence
of those we identify from perturbative calculations.

We emphasize again the necessarily ambiguous nature of the
magnitudes of the higher-twist matrix elements (e.g., (\ref{tildEop2})),
which depend on the manner in which the perturbative
integrals have been constructed and, indeed,
the order to which they have been computed.  In some sense, however,
the situation is a bit better than for the total
${\rm e}^+{\rm e}^-$ annihlation cross section, as described in
Ref.~\cite{mu2}, where beyond low orders, the perturbative series
will begin to show $\Gamma(n)$ behavior.  In resummation
formulas, such behavior is already included
with known coefficients.  If we succeed, therefore, in constructing
a perturbative exponent $S$ whose (arbitrary) higher
twist is not overly sensitive to higher-order corrections to
(for instance) $\Gamma_{\cusp}$, then the value of our nonperturbative
matrix elements will correspondingly be stable to higher-order
corrections in the perturbative calculation.  Of course, there
are other short-distance corrections not included in the
resummation, but these will appear only at still higher twist.

Our discussion has been quite formal, but
we hope that it will prove valuable to
recognize that a limited set of nonperturbative parameters
may enter into the first higher-twist corrections to
many high energy cross sections.  Such ``universality"
has proved of great practical use
when applied to local condensates in QCD sum rules,
despite the limitations and ambiguities inherent
in their combination with perturbative calculations.
Beyond this,
were it possible to actually compute, by some nonperturbative
method, highly nonlocal and relativistic operator combinations
such as eq.~(\ref{tildEop2}),  in a manner consistent with a
particular perturbative construction of $S$, then
the formalism presented here would provide new tests of the theory.

\hspace{2cm}

 {\em Acknowledgements \/}
 We are grateful for helpful conversations
 with  John Collins, Harry Contopanagos, Dave Soper,
 Al Mueller and Anatoly Radyushkin.
 This work was supported in part by the National
 Science Foundation under grant PHY 9309888 and by the Texas National
 Research Laboratory.

\end{document}